\documentclass[a4paper,fleqn]{cas-dc}

\usepackage[numbers,sort&compress]{natbib}
\usepackage{ulem}
\usepackage[version=4]{mhchem}
\usepackage{relsize}

\def\tsc#1{\csdef{#1}{\textsc{\lowercase{#1}}\xspace}}
\tsc{WGM}
\tsc{QE}
\tsc{EP}
\tsc{PMS}
\tsc{BEC}
\tsc{DE}

\begin{document}
\let\WriteBookmarks\relax
\def\floatpagepagefraction{1}
\def\textpagefraction{.001}

\shorttitle{Porous carbon nitride fullerenes}

\shortauthors{ZG Fthenakis}

\title [mode = title]{Porous carbon nitride fullerenes: a novel family of porous cage molecules}                      
\tnotemark[1]



%
\author[1,2,3]{Zacharias G. Fthenakis}[type=,
                        auid=,bioid=,
                        prefix=,
                        role=,
                        orcid=0000-0003-2661-5743]

\cormark[1]

\fnmark[1]

\ead{fthenak@eie.gr, zacharias.fthenakis@nano.cnr.it}

\ead[url]{https://www.nano.cnr.it/researcher-profile/zacharias-fthenakis/}


\affiliation[1]{organization={Theoretical and Physical Chemistry Institute, National Hellenic Research Foundation},
    addressline={}, 
    city={Athens},
    postcode={GR-11635}, 
    country={Greece}}
    
\affiliation[2]{organization={Istituto Nanoscienze-CNR},
    addressline={Piazza San Silvestro 12}, 
    city={Pisa},
    postcode={56127}, 
    country={Italy}}

\affiliation[3]{organization={NEST, Scuola Normale Superiore},
    addressline={Piazza San Silvestro 12}, 
    city={Pisa},
    postcode={56127}, 
    country={Italy}}

\author[1]{Nektarios N. Lathiotakis}[type=,
                        auid=,bioid=,
                        prefix=,
                        role=,
                        orcid=0000-0002-5589-7930]

\cormark[1]

\fnmark[2]

\ead{lathiot@eie.gr}

\ead[url]{http://www.eie.gr/nhrf/institutes/tpci/cvs/cv-lathiotakis-en.pdf}









\begin{abstract}
We propose and study theoretically a novel family of cage molecules, the porous carbon nitride fullerenes (PCNFs), 
which can be considered the zero-dimensional counterparts of the two-dimensional porous graphitic carbon nitrides, in accordance with icosahedral fullerenes, representing the zero-dimensional counterpart of graphene. We focus on two representative members of the PCNF family, the icosahedral \ce{C60N60} and \ce{C120N60} which are the first members of the two main sub-families of these structures derived from the class I and II Goldberg polyhedra, respectively. 
Taking into account that the two-dimensional graphitic carbon nitrides are considered the next-generation materials for several interesting applications, the interest in their zero-dimension counterparts is obvious.
In the present study, we utilize Density Functional Theory to find their structural, vibrational, and electronic properties. The stability of the proposed cages was demonstrated through vibrational analysis and molecular dynamics simulations with ReaxFF potentials. Regarding thermal stability, we found that \ce{C60N60} could be stable well above 1000~K and \ce{C120N60} well above 2000~K. Finally, we discuss several ideas regarding precursors that could be possibly used for their bottom-up synthesis.
\end{abstract}

\begin{graphicalabstract}
\includegraphics{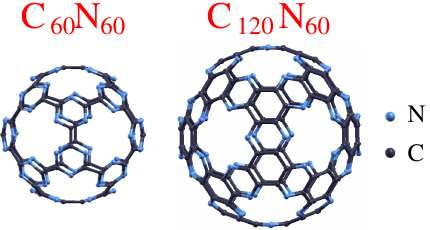}
\end{graphicalabstract}


\maketitle

\section{Introduction}

Given that fullerenes are the zero-dimensional (0D) analogues of graphene, a question that naturally arises is whether  0D fullerene analogues of two-dimensional (2D) porous graphitic carbon nitrides (g-C$_n$N$_m$) exist.

Porous g-C$_n$N$_m$ constitute a family of periodic graphene-based structures, incorporating substitutional N atoms in several sites of a porous graphene lattice\cite{Montigaud2000,doi:10.1021/ja809307s,doi:10.1021/jacs.9b00144,D0TA07437C,synth_C2N,doi:10.1021/acsnano.7b07473,https://doi.org/10.1002/adma.201501503,LI20061593,DaSilva2023}. A pore is formed in the absence of some C atoms of the pure graphene honeycomb lattice, and substitution of the two-fold coordinated C atoms at the pore boundary with N atoms (pyridinic N atoms). Pyridinic N atoms appear entirely at the pore edges, while three-fold coordinated N atoms (graphitic N atoms) may also appear in the structure, substituting three-fold coordinated C atoms. In these structures, N atoms are not favorable in nearest-neighbor lattice positions, since they are repulsive due to their negative charge \cite{PhysRevX.2.011003}. Thus, such an arrangement of N atoms does not appear in the structure.

Several structures in the porous g-C$_n$N$_m$ family have been reportedly synthesized. In the sub-family of structures containing both graphitic and pyridinic nitrogen atoms, representative examples are two distinct g-C$_3$N$_4$ structures, one consisting of triazine-based \cite{Montigaud2000} and one of heptazine-based \cite{doi:10.1021/ja809307s} units. These structures are predominantly observed in multilayer graphitic-like forms. Another heptazine-based structure of this sub-family, g-C$_3$N$_2$, is also reportedly synthesized. \cite{doi:10.1021/jacs.9b00144} 
 In the sub-family of structures with only pyridinic N atoms, notable examples are the g-\ce{C2N}  \cite{D0TA07437C, synth_C2N,doi:10.1021/acsnano.7b07473, https://doi.org/10.1002/adma.201501503, DaSilva2023} and g-\ce{CN} \cite{LI20061593, DaSilva2023}.
These structures have attracted considerable scientific interest over the last decade, exhibiting a broad spectrum of applications, which include catalysis \cite{CAO2021846,D1NJ05689A,C6CP03398A,C7MH00379J,D0CP00319K,catal13030578,ASHWINKISHORE201950,https://doi.org/10.1002/cphc.201600209,doi:10.1021/acs.jpcc.7b07776,doi:10.1021/ja809307s,D2CS00806H}, (such as photocatalytic water splitting \cite{ASHWINKISHORE201950,https://doi.org/10.1002/cphc.201600209,doi:10.1021/acs.jpcc.7b07776,doi:10.1021/ja809307s} and single atom catalysis  \cite{D2CS00806H}), gas separation \cite{RAO201953, C2N_1,C2N_2,C4RA15322G,C5TA05700K,CHANG2018294,WEI2024329} and nanofiltration \cite{D0CP02993A}, water desalination \cite{Mehrdad_2021}, hydrogen storage \cite{PANIGRAHI20217371}, and anode materials for Li-ion batteries \cite{https://doi.org/10.1002/adma.201702007}.
Given that the 0D counterparts of porous g-C$_n$N$_m$s will be structurally very similar to them, it is reasonable to expect that the above interesting properties of porous g-C$_n$N$_m$ would be inherited by their 0D counterparts. Furthermore, these 0D structures may exhibit additional novel properties due to their specific cage geometry and finite size, including properties arising from quantum confinement. Hence, the elevated interest in them is more than obvious.

Considering that (i) the relation between these 0D fullerene-like structures with porous g-C$_n$N$_m$ is analogous to the connection between icosahedral fullerenes and graphene, and (ii) the design of porous g-C$_n$N$_m$ involves the formation of pores and substitution of appropriate C atoms with N atoms, as described above, the form of these molecules can be easily derived from Goldberg polyhedra \cite{Goldberg}, in a similar way to the design of icosahedral fullerenes. The design of such 0D structures, which we call  ``{\it porous carbon nitride fullerenes}" (PCNFs), needs the formation of proper pores in icosahedral fullerenes in a way that retains full icosahedral symmetry. 

\begin{figure}[h!]
\includegraphics[width=0.47\textwidth]{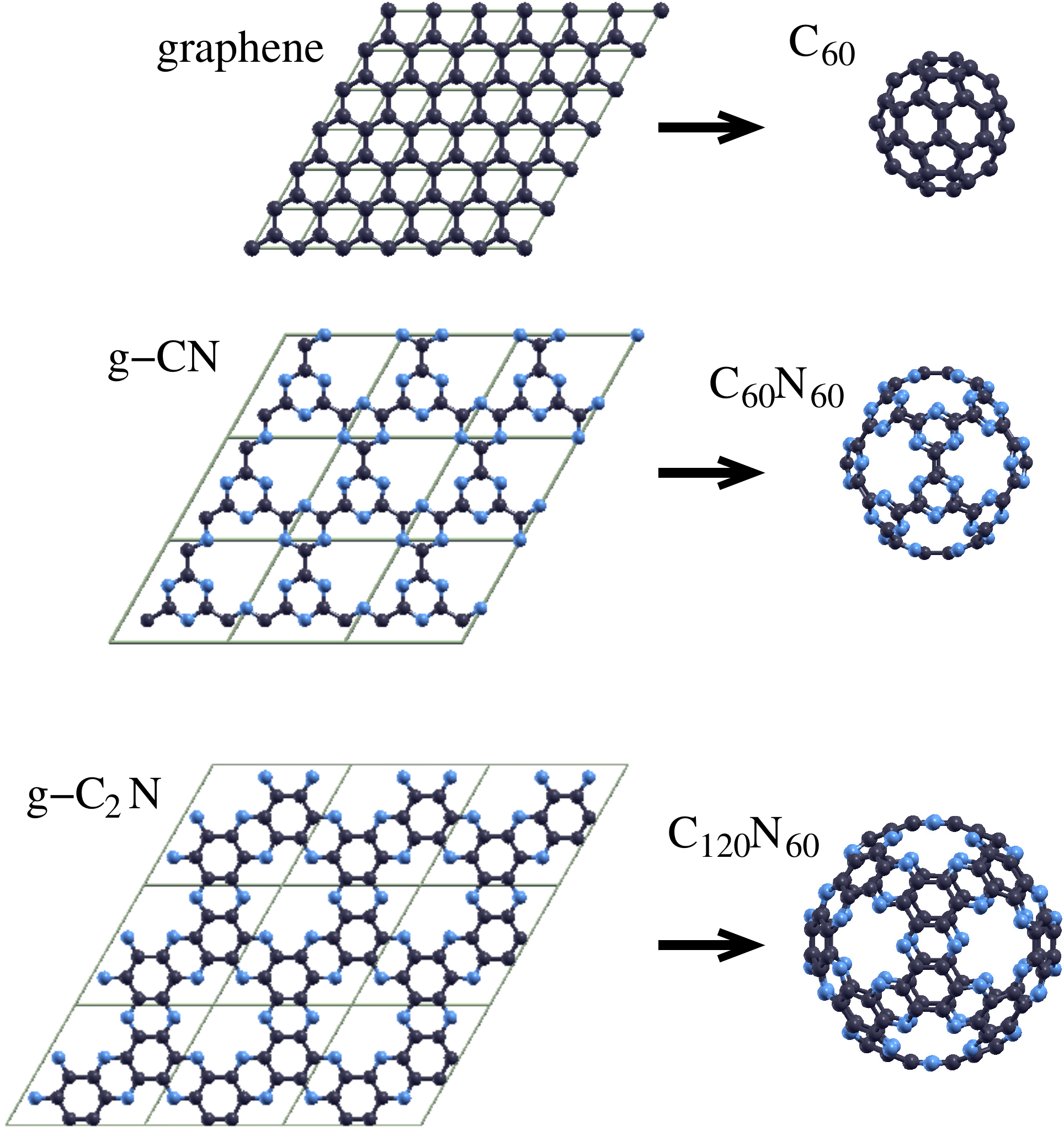} \\
\caption{\label{fig:full} 2D structures (graphene and g-C$_n$N$_m$s) and their fullerene counterparts.}
\end{figure}

Since only $(n,0)$ and $(n,n)$ Goldberg polyhedra have full icosahedral symmetry \cite{icos_ful}, the design of a fully icosahedral PCNF implies the introduction of several pores in the corresponding fullerene, by removing C atoms from the irreducible part of the triangles formed by the centers of neighboring pentagons in that fullerene. Then two-fold C atoms at the pore boundaries should be replaced by pyridinic N atoms. Maybe, other C atoms could be replaced by graphitic N atoms, as well, like in the design of g-C$_n$N$_m$s. It is easy to show that \ce{C20} and \ce{C80} fullerenes of the $(n,0)$ fullerene family (corresponding to the $(n,0)$ Goldberg polyhedra), and \ce{C60} fullerene of the $(n,n)$ fullerene family (corresponding to the $(n,n)$ Goldberg polyhedra) can not be the precursors in the design of PCNFs. As a consequence, the smallest members of the $(n,0)$ and $(n,n)$ fullerene families that can be the precursors of PCNF design are \ce{C180} and \ce{C240}, respectively. The PCNFs derived from them are the \ce{C60N60} and \ce{C120N60}, respectively, corresponding to the g-CN and g-\ce{C2N} 2D structures. 
These PCNFs (\ce{C60N60} and \ce{C120N60}) together with the corresponding 2D structures (g-CN and g-\ce{C2N}, respectively) are shown in Fig.~\ref{fig:full}. In the same figure, graphene and icosahedral \ce{C60} are also shown, to emphasize the relation between (i) graphene with \ce{C60}, (ii) graphene with g-C$_n$N$_m$s, and (iii) icosahedral fullerenes with PCNFs.

In the present work, we theoretically study \ce{C60N60} and \ce{C120N60} as representatives of the PCNF family. We investigate their stability (including their formation energies) and their structural, vibrational, and electronic properties. Our findings reveal that both structures are dynamically, thermally, and electronically stable. Following the presentation of our theoretical results, we discuss the potential for bottom-up synthesis of PCNFs using possible precursors.

\section{Methods}
For the structural, vibrational, and electronic properties,  we employ Density Functional Theory (DFT) with state-of-the-art functionals, namely the range-separated, hybrid $\omega$B97X-D~\cite{B810189B} and the hybrid meta-GGA MN15~\cite{C6SC00705H}, as implemented in the Gaussian 16 code \cite{g16}. For these calculations, the 6311-G(d,p) basis set was used, which was found appropriate for this study. The vibrational analysis certified that both PCNFs are dynamically stable structures. The ionization potentials (IPs) and electron affinities (EAs) were obtained as the absolute energy differences between the neutral molecule and positive and negative ions respectively. For the ions, separate calculations of the doublet states were performed, with full geometry relaxation. The large values of the obtained chemical hardness and HOMO-LUMO gaps demonstrate the stability of the electronic state of both PCNFs.

The formation energies of \ce{C60N60} and \ce{C120N60} were calculated using periodic DFT calculations with the PBE functional supplemented with Grimme's D3 dispersion corrections \cite{grimme}, as implemented in the Quantum Espresso code \cite{Giannozzi_2017}. This methodology allows the treatment of the two molecules and their periodic 2D counterparts at the same level of theory and orbital expansion. We adopted the projected augmented wave method (and pseudopotentials) and kinetic energy cutoffs for the wavefunctions and the density 60 and 500 Ryd, respectively. For the periodic structures, 24$\times$24$\times$1 k-mesh were used, while for the two molecules, $\Gamma$-point calculations were performed in a cubic supercell of a 36~\AA{} side. 

Regarding thermal stability, molecular dynamics (MD) simulations were performed under NVT conditions, for $0\le T\le 2000$~K, with a 100~K increment, using the CHON-2019 \cite{CHON2019} and the GR-RDX-2021 \cite{ReaxFFs} ReaxFF potentials. For these simulations, the LAMMPS \cite{LAMMPS} suite was employed, the NVT conditions were simulated using the Nos\'e-Hoover thermostat \cite{Nose,Hoover}, and the time step adopted was 0.25~fsec, for all cases. First, an annealing simulation was performed, under NVT conditions, using $2\times 10^6$ time steps, from 300~K down to 0~K, followed by a conjugate gradient optimization to find the energetically optimum structure. Before each NVT simulation at a specific temperature
$T$, thermalization of the optimized structure at that $T$ was performed, under NVT conditions, for $10^5$ time steps. Following thermalization, $2\times 10^6$ time steps of molecular dynamics simulation were performed, and the time average of the energy for each such simulation was calculated. Similar simulations were performed for \ce{C60} for comparison. These MD simulations reveal that \ce{C60N60} PCNF is stable well above 1000~K and \ce{C120N60} well above 2000~K.

\section{Results and discussion}
\subsection{Equilibrium geometry and structural features}
\begin{figure}[h!]
\centering
\includegraphics[width=0.98\columnwidth]{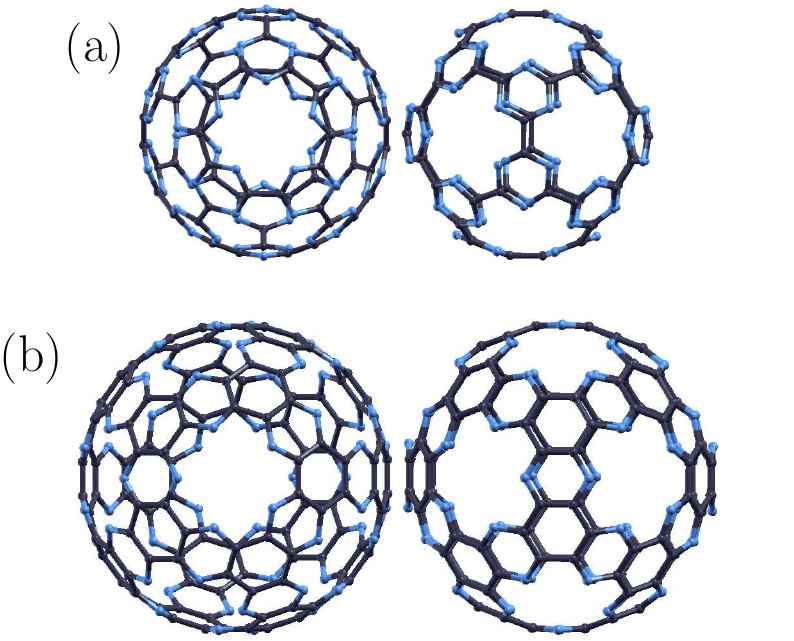}
\caption{\label{fig:struct} Structures of (a) \ce{C60N60} and (b) \ce{C120N60} from two different viewpoints. C atoms are shown in black and N atoms in blue. }
\end{figure}

In Fig.~\ref{fig:struct}, we show two different views for each of the equilibrium structures of \ce{C60N60} (panels (a)) and \ce{C120N60} (panel (b)). As we see, these two structures are composed of \ce{C3N3} and \ce{C4N2} hexagons, respectively, interconnected with C-C bonds. 
In \ce{C120N60},  these interconnecting bonds with the C-C bonds within \ce{C4N2} hexagons form hexagonal Carbon rings. The geometries of the structures are presented in the Supporting Information in Cartesian coordinates (xyz format). 

As one can see in Fig.~\ref{fig:full}, these structural features of \ce{C60N60} and \ce{C120N60}, as well as their stoichiometry, are identical with those in the g-\ce{CN} and g-\ce{C2N}, respectively.  The main morphological differences between the two PCNFs and their 2D counterparts are (i) the curvature of the PCNFs instead of the planarity of the g-CN and g-\ce{C2N}, and (ii) the pentagonal form of the pores in the two PCNFs instead of the hexagonal ones in the 2D structures. Considering an expansion of Fthenakis nomenclature scheme for graphene pores \cite{nomenclature1, nomenclature2}, to include PCNFs, their pores are labeled as 33333, while those of the g-\ce{CN} and g-\ce{C2N} as 333333. In comparison with the corresponding initial \ce{C180} and \ce{C240} C-fullerenes, the \ce{C60N60} and \ce{C120N60} PCNFs contain exclusively hexagonal rings, unlike C-fullerenes which contain also pentagons. Instead, they comprise those pentagonal 33333 pores.

The optimized geometries of the two PCNFs, obtained by our calculations with the $\omega$B97X-D functional, are shown in Fig.~\ref{fig:struct}. They are practically identical to those obtained with the MN15 functional, as seen in Tab.~\ref{tab:geom_c60n60}, where several structural properties of the two PCNFs are presented. 

The C-C and C-N bond lengths in $\ce{C60N60}$ were found to be $d_{C-C}\approx 1.52$~\AA{} and $d_{C-N}\approx 1.33$~\AA{}, respectively. The diameter of the pores, defined as the diameter of the circumscribed circle of the N-atoms pentagon is $D\approx 4.56$~\AA{}. The triazine hexagons \ce{C3N3} are almost planar, with only a small elevation, $\delta d=0.023$~\AA, of the N-atoms plane, outwards with respect to that of the C-atoms plane. Considering the \ce{C60N60} as a sphere, its diameter, defined as the distance between diametrically opposite C or N atoms, is $d_{C-C}^{sphere}\approx 11.3$~\AA{} and $d_{N-N}^{sphere}\approx 11.4$~\AA, respectively. 

For the \ce{C120N60}, two different kinds of C-C bonds exist: those interconnecting the \ce{C4N2} hexagons with length $d_{C-C}\approx 1.48$~\AA, and those of the \ce{C6N2} hexagons with lengths $d_{C-C}^\prime\approx 1.415$~\AA. The length of the C-N bonds is $d_{C-N}\approx 1.33$~\AA, as in the case of \ce{C60N60}. The N-atoms elevation outwards is $\delta d\approx 0.66$~\AA{}, almost three times larger than that of \ce{C60N60}, although still small. The diameter $D$ of the pores is $D\approx 4.565$~\AA, i.e. practically the same as that of \ce{C60N60} pores. As expected, the corresponding sphere diameter is larger than that of \ce{C60N60} by approximately 2~\AA, with values  $d_{C-C}^{sphere}\approx 13.3$~\AA{} and $d_{N-N}^{sphere}\approx 13.5$~\AA.

For comparison, we performed optimization calculations for the icosahedral \ce{C60} fullerene with the same methods. According to our findings, the length of the common bonds of adjacent pentagons and hexagons (single bonds) is $d_{hp}\approx 1.45$~\AA{}, and the length of the common bonds of adjacent hexagons (double bonds) is $d_{hh}\approx 1.39$~\AA. These values are comparable with the $d_{C-C}$ and $d_{C-C}^\prime$ bond lengths of \ce{C60N60} and \ce{C120N60}, with $d_{C-C}^{\textrm{\ce{C60N60}}} > d_{C-C}^{\textrm{\ce{C120N60}}}> d_{hp}$ and $d_{C-C}^\prime > d_{hh}$. This is a strong indication that the C-C bonds interconnecting \ce{C3N3} hexagons in \ce{C60N60} and \ce{C4N2} hexagons in \ce{C120N60} exhibit single bond character. Instead, those of the \ce{C4N2} hexagons have the character of a double bond. Moreover, it indicates that the C-C bonds of \ce{C60N60} and \ce{C120N60} are weaker than those of \ce{C60}, with those of \ce{C120N60} being stronger than those of \ce{C60N60}.

Since the diameter $D$ and the atomic environment of the pores for both \ce{C60N60} and \ce{C120N60}  are practically the same, the molecular permeation properties of those pores are expected to be very similar. The local curvature is the only parameter that might differentiate these properties, varying between the two PCNFs due to their different sphere diameters. It is worth noting, that the size of the pore diameter is within the range of interesting pores for the permeation of small molecules \cite{RAO201953,C5TA05700K,C2N_1,C2N_2,WEI2024329}.
 We should mention that several members of the PCNF family, incorporate these pentagonal pores, which (to the best of our knowledge) have not been observed in any other structure, so far.  On the other hand, such pentagonal pores might also exist as defects in g-C$_n$N$_m$s, modifying the structure and impairing its planarity locally. The properties related to such pores make these structures particularly interesting.

\begin{table}[h!]
\begin{tabular}{l@{\hskip7pt}c@{\hskip3pt}c@{\hskip7pt}c@{\hskip3pt}c}
& \multicolumn{2}{c}{\ce{C60N60}} & \multicolumn{2}{c}{\ce{C120N60}} \\
\small{Functional} & \footnotesize{$\omega$B97X-D}\! & \footnotesize{MN15} & \footnotesize{$\omega$B97X-D}\!  & \footnotesize{MN15} \\
\hline
$d_{C-C}$  & 1.520 & 1.518 & 1.478 & 1.477\\
$d_{C-C}^\prime$  &  & & 1.413 & 1.417 \\
$d_{C-N}$  & 1.327 & 1.330 & 1.325 & 1.328 \\
$\delta d$  & 0.023 & 0.023 & 0.066 & 0.065 \\
$D$  & 4.553& 4.561 & 4.561 & 4.573 \\
$d^{sphere}_{C-C}$  & 11.314 & 11.332 & 13.314 & 13.321 \\
$d^{sphere}_{N-N}$  & 11.392 & 11.410 & 13.461 & 13.465 \\
\hline
IP & 9.20 & 9.23 & 8.56 & 8.08 \\
EA & 2.08 & 2.60 & 2.19 & 2.79 \\
$\eta$ & 3.56 & 3.32 & 3.18 & 2.64 \\
HOMO & -9.73 & -8.68 & -8.65 & -7.61 \\
LUMO & -2.02 & -3.15 & -2.23 & -3.28 \\
H-L Gap & 7.71 & 5.53 & 6.42 & 4.32 \\
\multicolumn{5}{c}{Mulliken Population} \\
 C atoms & 0.24 & 0.20 & 0.12 & 0.095 \\
 N atoms & -0.24 & -0.20 & -0.24 & -0.19 \\
 \hline
 min. $\nu$ & 75.6 & 76.1 & 90.0 & 88.2 \\
\end{tabular}
\caption{\label{tab:geom_c60n60} Structural, electronic and vibrational features of \ce{C60N60} and \ce{C120N60}  calculated utilizing the $\omega$B97X-D and MN15 functionals and the 6-311G(d,p) basis set. All interatomic distances, energies, charges, and minimum frequencies (min. $\nu$) are given in \AA, eV, electrons, and cm$^{-1}$, respectively.}
\end{table}

Molecular cage structures, like PCNFs, hold promise as traps for small molecules or metal clusters. Given the size of the radius of the PCNFs, one can expect that up to a few small molecules can be accommodated in these cages. Molecule trapping in these systems has the potential for a wide range of applications.

\subsection{Electronic properties}
Due to the icosahedral symmetry of the \ce{C60N60} and \ce{C120N60}, the energy levels are in most cases degenerate. In particular, the HOMOs for the \ce{C60N60} and the \ce{C120N60} are five- and three-fold degenerate, respectively, while the LUMOs are three- and five-fold degenerate, respectively. However, in both cases, the HOMO orbitals are fully occupied, enhancing stability, like the aromatic ring molecules. 

\begin{figure}[h!]
\centering
\includegraphics[width=0.98\columnwidth]{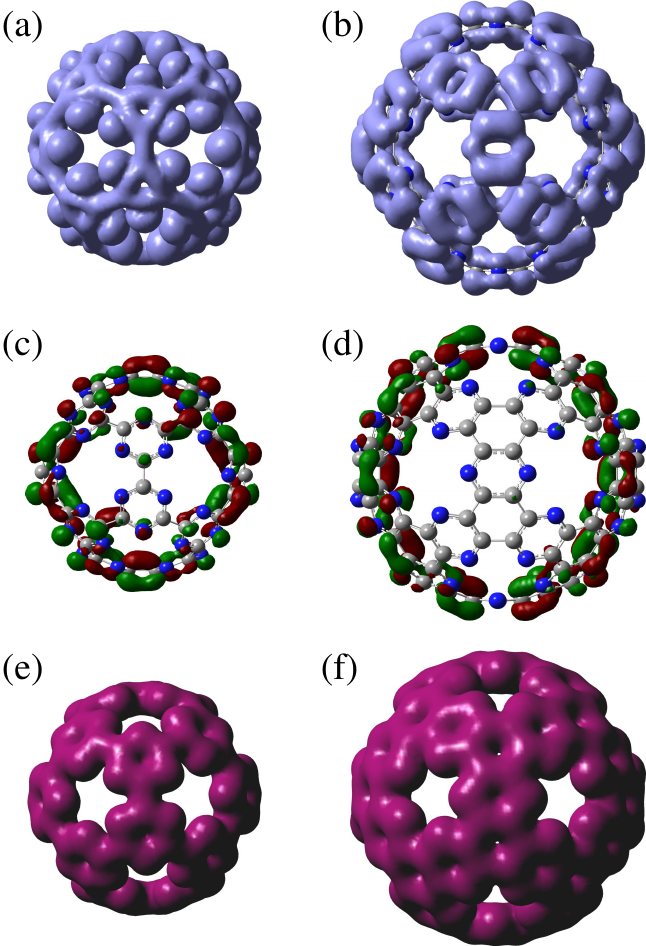}
\caption{\label{fig:el_dens} Electron density corresponding to all degenerate HOMOs of \ce{C60N60} (a) and  \ce{C120N60} (b). One of the degenerate LUMOs of \ce{C60N60} (c) and  \ce{C120N60} (d). Total electron density of \ce{C60N60} (e) and  \ce{C120N60} (f). }
\end{figure}

The electron density corresponding to all five (for the \ce{C60N60}) and all three (for the \ce{C120N60}) degenerate HOMOs are shown in Fig.~\ref{fig:el_dens}(a) and (b), respectively. As we see, they have entirely different characters. The HOMOs of \ce{C60N60} are dominated by Nitrogen sp$^2$ lone-pair states, while for \ce{C120N60}, they have the character of p$_\perp$ orbitals. With p$_\perp$ we mean p-orbitals directed perpendicular to the \ce{C120N60} surface locally. Moreover, one can see that the HOMOs electron density for \ce{C120N60} is high along the C-C bonds of the \ce{C4N2} hexagons, indicating the presence of a double bond. Instead, along the C-C bonds interconnecting the \ce{C4N2} hexagons, it almost vanishes, in consistency with the presence of a single bond.  We previously discussed the nature of these bonds concerning their length and the electron density plots confirm it. One can also see in Fig.~\ref{fig:el_dens}(b) that the p$_\perp$ contribution of N atoms is relatively small compared to that of the C atoms.

The LUMOs, on the other hand, are of p$_\perp$ orbital character for both structures, as shown in Fig.~\ref{fig:el_dens}(c) and (f), which depict one of the degenerate LUMOs of \ce{C60N60} and \ce{C120N60}, respectively. Although not shown here, the other degenerate LUMOs have the same character differing mainly in their location. In Figs.~\ref{fig:el_dens}, we show the total electron density of \ce{C60N60} and \ce{C120N60} for completeness.

Our findings on the nature of the HOMOs' and LUMOs' of \ce{C60N60} and \ce{C120N60} are very similar to those for the valence and conduction bands of g-\ce{CN} and g-\ce{C2N}, respectively.  For g-CN, the valence band consists of the sp² orbitals of the nitrogen atoms' lone-pair electrons, lying in the structure's plane and localized on the nitrogen atoms. Instead, the conduction band is composed of p$_z$ orbitals of the C and N atoms \cite{C4RA15322G}. For the g-\ce{C2N}, the valence band is composed mainly of the p$_z$ orbitals of the C atoms, with a smaller contribution from the p$_z$ orbitals of the N atoms. Instead, the conduction band is composed of p$_z$ orbitals, of the N atoms mainly \cite{doi:10.1021/acs.jpcc.7b07776}.

For both structures, there is a relatively large HOMO - LUMO gap (H-L Gap) in comparison to the differences of the other successive energy levels of the occupied and unoccupied (virtual) orbitals, indicating a clear splitting between the $\pi$ occupied states from the $\pi^*$ unoccupied ones. The calculated HOMO, LUMO, and H-L gap values are shown in Tab.~\ref{tab:geom_c60n60}, together with the IP, EA, and chemical hardness, $\eta=({\rm IP}-{\rm EA})/2$. Notably, the HOMO and LUMO absolute values are good measures of IPs and EAs for the state-of-the-art approximations chosen. The large values of H-L gap  and $\eta$ strongly suggest that the electronic state of the studied PCNF structures is particularly stable. It is worth mentioning that,  the EA values of PCNFs are large, like that of \ce{C60}, hence they could be promising candidates as electron acceptors in several applications. The values of the IPs are also exceptionally large leading to large $\eta$. 

For comparison, we also calculated the IP, EA, and H-L gap of \ce{C60} with $\omega$B97X-D and MN15 functionals. We obtained IP=$8.56$, $8.01$~eV, EA=$2.32$, $2.45$~eV, and H-L Gap = $6.00$, $3.92$~eV, respectively. Thus, the chemical hardness for PCNFs is larger than that of \ce{C60} which is 2.84, 2.67 eV for $\omega$B97X-D and MN15, respectively. The larger values of the H-L Gap and chemical hardness, compared to \ce{C60}, indicate that \ce{C60N60} and \ce{C120N60} are electronically more stable. 
The drop in the HOMO-LUMO gap from \ce{C60N60} to \ce{C120N60} seems to be similar to the behavior of the gap of graphene nanoribbons, which decreases as their width increases \cite{PhysRevLett.97.216803}.

Furthermore, Mulliken population analysis reveals a weak charge transfer from N atoms to C atoms in both structures. Specifically, about 0.2 electrons move from each N atom and are equally distributed to the C atoms. Despite the charge transfer in the C-N bonds, the total dipole moment of PCNFs is zero, due to their icosahedral symmetry.

Regarding vibrational analysis, the results show that all eigenvalues of the dynamical matrix are positive (real vibrational frequencies), indicating dynamic stability. The minimum frequencies obtained are shown in Tab.~\ref{tab:geom_c60n60} ($\nu_{min}\approx 76$~cm$^{-1}$ for \ce{C60N60} and $\nu_{min}\approx 89$~cm$^{-1}$ for \ce{C120N60}). The corresponding minimum frequency value for \ce{C60} is $\nu_{min}\approx 266$~cm$^{-1}$, indicating that \ce{C60} is dynamically more stable. This might be one of the reasons that according to MD simulations (see below), the fracture of PCNFs occurs at lower temperatures than \ce{C60}.

\subsection{Formation energies of PCNFs}
With the periodic DFT calculations described above, we calculated the formation energy per atom, $\Delta E^{(f)}$, of \ce{C60N60}, \ce{C120N60} with respect to their 2D counterparts g-CN, g-\ce{C2N}, respectively. For comparison, we also calculated $\Delta E^{(f)}$ of  Ih-\ce{C60} fullerene with respect to graphene. Using the calculated total energy 
$E_{\rm PCNF}$ of the PCNF and the total energy $E_{\rm 2D}$ of its 2D counterpart per unit cell, 
\begin{equation}
\Delta E^{(f)} = \frac{E_{\rm PCNF}}{N_{\rm PCNF}} - \frac{E_{\rm 2D}}{N_{\rm 2D}}\,,
\end{equation}
where $N_{\rm PCNF}$, is the number of atoms of the PCNF and $N_{\rm 2D}$ is the number of atoms in the unit cell of the corresponding 2D structure.
We found that $\Delta E^{(f)}_{\ce{C60N60}}=0.09$~eV/atom and $\Delta E^{(f)}_{\ce{C120N60}}=0.08$~eV/atom, while 
$\Delta E^{(f)}_{\ce{C60}}=0.38$~eV/atom. 
The small values of $\Delta E^{(f)}$ of \ce{C60N60} and \ce{C120N60} in comparison with that of \ce{C60} can be explained in terms of the large bond length deviations of \ce{C60} from the bond length of graphene ($d=1.42$~\AA), and the corresponding relatively large bond angle deviations of the pentagonal angles (108$^o$) from the hexagonal ones of graphene (120$^o$). The absence of pentagonal rings in \ce{C60N60} and \ce{C120N60} and the insignificant deviations of their bond lengths from the corresponding ones of g-CN and g\ce{C2N}, respectively, significantly reduce the additional energy to form the PCNFs, which is mainly attributed to the strain energy due to the curvature.

For completeness, we calculated the formation energies, $E^{(f)}$, of 
g-CN, g-\ce{C2N}, \ce{C60N60}, \ce{C120N60}, and Ih-\ce{C60} fullerene with respect to graphene and \ce{N2} molecule, defined as
\begin{equation}
E^{(f)} = \frac{1}{N} \left[ E - n_{\rm C} \,\mathlarger{\epsilon}_{\rm g} -n_{\rm N} \,\mathlarger{\epsilon}_{{\rm N}_2} \right] \,
\end{equation}
where $N$ is the number of atoms and $E$ total energy of the PCNF, while $n_{\rm C}$, $n_{\rm N}$ are the numbers of C and N atoms respectively, and $\mathlarger{\epsilon}_{\rm g}$, $\mathlarger{\epsilon}_{{\rm N}_2}$ the total energies per atom of graphene and \ce{N2}, respectively. The formation energies we obtained are listed in Tab.~\ref{tab:formE}. As we see, the $E^{(f)}$ of \ce{C60N60} is approximately the same as that of \ce{C60}, while that of \ce{C120N60} is smaller. Also, that of g-\ce{C2N} is smaller than that of g-CN, which is unsurprising as this structure is ``closer'' to graphene with a larger density of hexagonal rings and less percentage of empty space. 

\begin{table}[h!]
     \begin{tabular}{c|ccc}
          0D & \ce{C60N60} & \ce{C120N60} & \ce{C60} \\
     $E^{(f)}$ (eV) & 0.38 & 0.31 & 0.38 \\ \hline
     2D & g-CN & g-\ce{C2N} & graphene \\
      $E^{(f)}$ (eV) & 0.29 & 0.23 & 0.00
          \end{tabular}
\caption{\label{tab:formE} Calculated formation energies,  $E^{(f)}$ with respect to graphene and \ce{N2} molecule.}
\end{table}

\subsection{Thermal stability}

\begin{figure}[h!]
\includegraphics[width=0.47\textwidth]{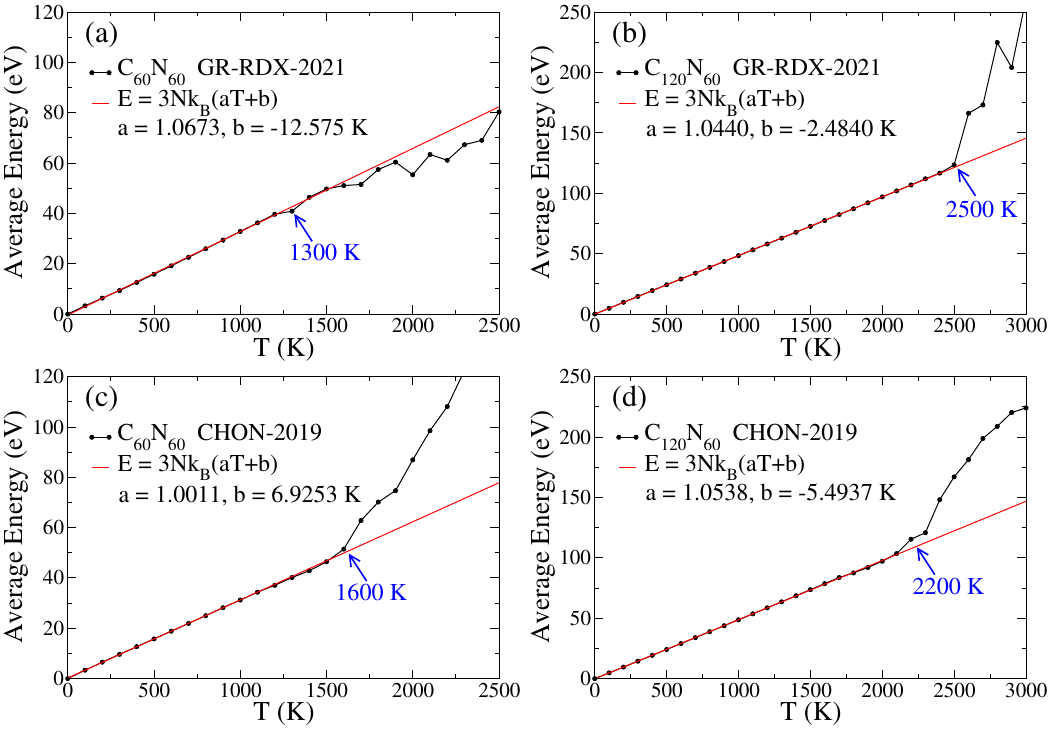} \\
\caption{\label{fig:vs_T} Average energy vs temperature obtained from MD simulations (i) using GR-RDX-2021 ReaxFF for \ce{C60N60} (a) and \ce{C120N60} (b), and (ii) using CHON-2019 ReaxFF for \ce{C60N60} (c) and \ce{C120N60} (d). For \ce{C60N60}, $N=120$ and for \ce{C120N60}, $N=180$. $k_B=1/11604.5$~eV/K. The predicted temperatures at which the transition takes place are shown in blue.}
\end{figure}

Figure~\ref{fig:vs_T} shows the time average of the energy $E$ as a function of temperature $T$ obtained from the MD simulations using the CHON-2019 and GR-RDX-2021 ReaxFFs for the \ce{C60N60} (left panels) and \ce{C120N60} (right panels). The linear relation between $E$ and $T$ at low temperatures represents the Dulong and Petit law, $E=3Nk_BT$, which indicates the existence of a single phase. That phase is the ``solid'' state of the molecule. In the temperature range where this relation is valid, only atomic vibrations near the equilibrium positions occur. A divergence in the slope ($3Nk_B$) of the Dulong and Petit law indicates the initiation of a phase transition, usually accompanied by the coexistence of more phases. This phase transition finally leads to the fracture of the molecule. As shown in Fig.~\ref{fig:vs_T}, the transition temperature for the \ce{C60N60} is well above 1000~K (1300~K for GR-RDX-2021 and 1600~K for CHON-2019) and for \ce{C120N60} well above 2000~K (2500~K for the GR-RDX-2021 and 2200~K for the CHON-2019). For comparison, we performed similar MD simulations for \ce{C60} with CHON-2019, and we found that the transition starts at around $3800$ K. This value is in agreement with the work of Kim and Tom\'anek \cite{PhysRevLett.72.2418}, reporting a transition temperature between 3400 and 3800~K using tight-binding MD simulations. While the estimated transition temperatures and minimum vibration frequencies for \ce{C60N60} and \ce{C120N60} are lower than those of C$_{60}$, they are still sufficiently high, to guarantee that both \ce{C60N60} and \ce{C120N60} are thermally stable at temperatures exceeding substantially 1000~K and 2000~K, respectively. 

\subsection{On the feasibility of synthesizing \ce{C60N60} and \ce{C120N60}}
Although we found that both \ce{C60N60} and \ce{C120N60} are stable even at temperatures well above 1000~K and 2000~K, respectively, the most important issue is the existence of efficient routes for their synthesis. Likely, the most promising route involves bottom-up techniques, using proper precursors, similar to the synthesis of g-C$_n$N$_m$s.  
Moreover, with proper nanosphere precursors including nitrogen functional groups, nitrogen-doped carbon nanospheres including pyridinic N atoms have already been synthesized \cite{doi:10.1021/acsanm.3c04289}.
Additionally, the synthesis of carbon nanospheres with g-\ce{C3N4} pores have also been reported \cite{doi:10.1021/acs.inorgchem.9b01089,TANG2020458} as well as the synthesis of spherical carbon nitride nanostructures \cite{doi:10.1021/nl015626h} with diameters ranging between 30~nm and 20~$\mu$m. In similar experimental studies, the formation of hollow mesoporous carbon nitride microspheres has also been reported \cite{LI2020186}. 
Therefore, the reported synthesis of both 2D g-C$_n$N$_m$ membranes \cite{D0TA07437C,C7CP02711G,D2CS00806H,WANG2022482,DaSilva2023} and carbon nanospheres  \cite{https://doi.org/10.1002/anie.201102011,D3CC00402C} as well as the minimal difference in the calculated formation energies of PCNFs compared to their 2D counterparts provide strong evidence for the feasibility of PCNF synthesis.

A possible precursor for the synthesis of \ce{C60N60} could be the 1,3,5-Triazinane (\ce{C3H9N3}), or the 1,3,5-Triazine, also called s-Triazine (\ce{C3H3N3}), or similar structures, like melamine (\ce{C3H6N6}) or cyanuric chloride (\ce{C3N3Cl3}). Cyanuric chloride has been involved (a) together with melamine in the synthesis of 2D g-\ce{C3N4} \cite{CNs}, (b) in the synthesis of carbon nitrogen nanotubes \cite{TRAGL2007529}, as well as (c) the formation of CN nanotube bundles, nanoribbons, and microspheres \cite{LI20061593} with 1:1 stoichiometry \cite{doi:10.1021/cm0493039}.
It is worth noting that one of the synthetic routes of the 2D g-\ce{C3N4} uses Cyanuric Chloride and melamine \cite{CNs}. Moreover, the synthesis of the three star-shaped 1,3,5-Triazine derivatives \cite{B913423A} is also very promising.

A possible precursor for the synthesis of \ce{C120N60} could be (i) either Pyrazine (\ce{C4H4N2}), or 1,2-Dihydropyrazine (\ce{C4H4(NH)2}), or similar molecules which include the \ce{C4N2} hexagon, or (ii) 1,3,5-Benzenetriamine (\ce{C6H3(NH2)3}), or 1,3,5-Trinitrobenzene (\ce{C6H3(NO2)3}), or similar molecules, which include the C$_6$ hexagon with three bonded N atoms. The synthesis of Pyrazinacenes \cite{C1OB05454F} based on Pyrazine may be a good starting point. Moreover, it is very promising for the synthesis of \ce{C120N60} the synthesis of its 2D counterpart, g-\ce{C2N}, via (a) a bottom-up wet-chemical reaction, using the precursors hexaaminobenzene trihydrochloride and hexaketocyclohexane octahydrate \cite{synth_C2N} and (b) polymerization of hexaaminobenzene (precursor), produced using amination of chloroanilic acid \cite{doi:10.1021/acsnano.7b07473}, (c) eutectic syntheses using as precursors the cyclohexanehexone and urea \cite{https://doi.org/10.1002/adma.201501503}. Other synthetic methods for g-\ce{C2N} have been also reported \cite{D0TA07437C}.

We should note that this work is purely theoretical and the discussion in the present section provides simply some arguments supporting the possibility of PCNF synthesis.  Synthesis, however, is definitely beyond the scope of the present work.

\section{Conclusions}
In this study, we present a novel family of structures, the porous carbon nitride fullerenes, representing the zero-dimensional analog of the two-dimensional porous carbon nitrides, just as fullerenes can be seen as the zero-dimensional counterparts of graphene. Our study focuses on the first members of that family of structures with icosahedral symmetry, (namely the \ce{C60N60} and \ce{C120N60}), which are derived from the class I and II Goldberg polyhedra, respectively, following the two $C_N$ icosahedral fullerene subfamilies with $N=20n^2$, (\ce{C20} subfamily) and  $N=60n^2$, (\ce{C60} subfamily), $n=1,2,3,\ldots$, respectively.

We performed DFT calculations with two state-of-the-art functionals ($\omega$B97X-D and MN15) and determined the optimal geometrical properties, the HOMO and LUMO energies and the vibrational frequencies of  
\ce{C60N60} and \ce{C120N60}. We also determined their  IPs and EAs as absolute total energy differences of the neutral structures from the positive and negative ions, respectively.  We found that  \ce{C60N60} and \ce{C120N60} are dynamically stable, while the large values of the H-L gap and chemical hardness indicate a robust electronic state. Due to their large EAs, these molecules could be used as electron acceptors.
The two structures have 12 identical N-terminated pentagonal-like pores each, with practically equal pore diameter ($D\approx 4.56$~\AA), which is in the diameter range of the interesting pores for the permeation of small molecules and they can capture metallic atoms for catalytic applications. The unique features of the pentagonal-like pores and the particular features rising from the cage shape of those molecules make them potentially interesting for several applications well beyond the applications of their 2D counterparts. 

Moreover, we performed molecular dynamics simulations under NVT conditions with the CHON-2019 and GR-RDX-2021 ReaxFFs, and we showed that the \ce{C60N60} and \ce{C120N60} PCNFs are thermally stable well above 1000~K and 2000~K, respectively. Additionally, we discuss possible precursors, which could be used in a bottom-up synthesis of these structures. We expect the new proposed molecules to be of great importance after their synthesis, as porous molecules, in trapping small molecules, and in catalysis.

\section*{Acknowledgments}
ZGF acknowledges financial support from the project PRIN 2022 - Cod. 202278NHAM (PE11) CHERICH-C ``Chemical and electrochemical energy storage materials from organic wastes: the treasure hidden in C-based materials'' - CUP B53D23008590006, funded by the European Union – Next Generation EU in the context of the Italian National Recovery and Resilience Plan, Mission 4, Component 2, Investment 1.1, ``Fondo per il Programma Nazionale di Ricerca e Progetti di Rilevante Interesse Nazionale (PRIN)''. 

NNL acknowledges support by the framework of the Action ``Flagship Research Projects in challenging interdisciplinary sectors with practical applications in Greek Industry'', implemented through the National Recovery and Resilience Plan Greece 2.0 and funded by the European Union - NextGenerationEU (Acronym: 3GPV-4INDUSTRY, project code: TAEDR-0537347).


\bibliographystyle{unsrtnat}
\bibliography{PCNFs.bib}

\end{document}